\documentclass[prb,amsmath,twocolumn,showpacs,superscriptaddress]{revtex4}
\usepackage{color}
\usepackage{graphics}
\usepackage{subfigure}
\usepackage{ulem}
\usepackage{times}
\usepackage{graphicx}
\usepackage{amsmath}
\usepackage{amsbsy}
\usepackage{dcolumn}
\usepackage{bm}
\usepackage{epsfig}
\usepackage{color}
\usepackage{mathtools}
\usepackage{dsfont}

\begin{document}

\title{Elastic coupling and spin-driven nematicity in iron-based superconductors}

\author{U. Karahasanovic}

\affiliation{Institut f\"ur Theorie der Kondensierten Materie, Karlsruher Institut
f\"ur Technologie, DE-76131 Karlsruhe, Germany}

\affiliation{Institut f\"ur Festk\"orperphysik, Karlsruher Institut f\"ur Technologie,
DE-76131 Karlsruhe, Germany}

\author{J. Schmalian}

\affiliation{Institut f\"ur Theorie der Kondensierten Materie, Karlsruher Institut
f\"ur Technologie, DE-76131 Karlsruhe, Germany}

\affiliation{Institut f\"ur Festk\"orperphysik, Karlsruher Institut f\"ur Technologie,
DE-76131 Karlsruhe, Germany}

\date{\today}
\begin{abstract}
Spin-driven nematic order that has been proposed for iron-based superconductors is generated
by pronounced fluctuations of a striped density wave state. On the other hand it is a well known fact that nematic order parameter
couples bilinearly to the strain, which supresses the fluctuations of the nematic order parameter itself and lowers the upper critical
dimension, yielding mean-field behaviour of the nematic degrees of freedom for $d>2$. This is consistent with the measured Currie-Weiss behaviour of the nematic susceptibility. Here we reconcile this apparent contradiction between pronounced magnetic fluctuations and mean-field behaviour of the nematic degrees of freedom. We show, by developing a $\varphi^4$ theory for the nematic degrees of freedom, that the coupling to elastic strain does not suppress the fluctuations that cause the nematic order in the first place (magnetic fluctuations), yet it does transform the Ising-nematic transition into a mean-field transition. In addition, we demonstrate that the mean field behavior occurs  in the entire temperature regime where a softening of the shear modulus is observed.
\end{abstract}


\maketitle

\section{Introduction}

\label{intro}

In many iron based superconductors a structural phase transition that sets at in at the temperature $T_s$, from the  high-temperature tetragonal phase into an orthorhombic phase, has been shown to closely track the magnetic transition at $T_m$ \cite{Birgeneau11, Kim11, matsuda_t, nematic_review}, i.e.: $T_s\geq T_m$. 
In $1111$ materials the structural and magnetic phase transitions are split and of second order. On the other hand in the $122$ family, the transitions are either joint and of first order or split and second order. In Ba(Fe$_{1-x}$Co$_{x}$)$_{2}$As$_{2}$, where the transitions are split,  the lower, magnetic transition is weakly first order for $x < 0.02$.\cite{Kim11,Birgeneau11} 

From elastoresistance measurements\cite{Chu10} there is very strong evidence that the nematic state is driven by electronic excitations. This is fully consistent with the comparatively large resistivity-anisotropy measurements \cite{Tanatar10,Chu2012} with the softening of the elastic modulus over a wide temperature range\cite{Fernandes13_shear,Fernandes2010,Kontani1,Kontani2,Anna1}, with strong signatures in the electronic Raman response in the normal, \cite{Gallais,Rudi,Blumberg1,Blumberg2,Kontani1,Khodas15,Una,yann_review} 
 and superconducting\cite{Gallais15} states,  and with
the observation of anisotropies in various additional observables,
such as thermopower, \cite{Jiang2013} optical conductivity, \cite{Dusza2011,Nakajima2011}
torque magnetometry, \cite{matsuda_t} and in STM measurements.\cite{Rosenthal13}

To explain the origin of the nematic phase in pnictides  an orbital fluctuation based scenario,\cite{Phillips11,Applegate11,Dagotto13,w_ku10,kruger1,kruger2}
and a theory for spin-driven nematicity \cite{Fernandes12,naturereview,nematic_review} have been proposed.
In the latter, spin-fluctuations, associated with striped magnetic order, can generate the emergent electronic nematic order at temperatures above the Neel temperature.\cite{Fernandes12,naturereview,nematic_review,Xu08,Fang08}
Nematic degrees of freedom couple to the lattice\cite{Qi09,Fernandes2010,Cano10} and induce
the structural phase transition to the ortorhombic phase.
Scaling of the shear modulus and the NMR spin-lattice relaxation rate for 122 systems, discussed in Ref. \onlinecite{Fernandes13_shear}, strongly support the spin-driven nematicity scenario. Here the presence of the magnetic fluctuations associated with the stripe density wave phase proves crucial for stabilizing the nematic phase. On the other hand, nematic
order was also observed in FeSe, a material where spin-magnetic order is only  generated via application of external pressure, while a structural transition along with a softening of the lattice occurs at ambient pressure in a fashion very similar to the 122 iron pictides.\cite{Terashima15, Anna15, Baek15} This led to the suggestion that this system might be driven by orbital fluctuations. Alternatively, given  the observed small Fermi surfaces of FeSe \cite{Watson15, Terashima14} almost degenerate imaginary charge density and spin density waves are expected.\cite{Chubukov15,RG1, RG2, RG3} Therfore, a mechanism based on fluctuating and possibly ordered striped imaginary
charge density waves was proposed that is conceptually very similar
to the striped spin-driven mechanism.\cite{Chubukov15}

In Ref. \onlinecite{Fernandes12} the phase diagram of striped density
wave induced nematicity was investigated. At a finite temperature
phase transition, it was found that pre-emtive nematic order emerges
via split phase transitions for some regime of the parameter space
of the model, while in other cases a joint first order transition
occurs, all in good agreement with experiment.  Clearly,  fluctuations were crucial to derive this rich phase diagram.

On the other hand: the behavior near the nematic transition seems to display generic mean field behavior, including the Currie-Weiss behaviour of nematic susceptibility. \cite{Anna1, Anna3, yoshizawa} In addition, 
 it was noted already in 1970s, \cite{Levanyuk70,Cowley76,Folk76,Folk79} that fluctuations of an  order parameter that couples
linearly to an elastic deformation are suppressed. This is
the case for the nematic order parameter which couples linearly to
orthorombic distortion via the nemato-elastic coupling.  This can lead to mean-field behaviour, instead of $d=2$ or $d=3$ Ising like behavior
that is expected in the absence of the coupling to strain.\cite{Levanyuk70,Cowley76,Folk76}
In the context of the iron based systems, this effect was already stressed by Cano et al., \cite{Cano10} and in Ref. \onlinecite{Zacharias15},
where quantum critical elasticity was investigated. Given these observations and the fact that fluctuations
were essential for the derivation of the aforementioned phase diagram, 
it seems worthwile exploring whether the strain coupling induced mean
field behavior of the nematic degrees of freedom can change the conclusions of Ref. \onlinecite{Fernandes12}. 
 
To address this problem we start from a model of a spin-driven nematic phase, similar to the Ref. \onlinecite{Fernandes12} and include the coupling to elastic strain.
We then determine an effective order parameter theory of the nematic
order parameter in the presence of strain. Integrating out strain
results in the appearance of the non-analytic terms in the propagator
for nematic fluctuations, that are directional dependent, similar
to what was found in Refs. \onlinecite{Aharony73, Aharony73B, Chalker80}, where spin models in the presence of dipolar interactions were examined.
These non-analytic terms lead to mean field behavior of the nematic
transition and are shown to give rise to a Curie Weiss susceptibility over a sizable temperature range. Yet, the strain coupling does not affect  the very existence of nematic order
and of the split structural / magnetic transition temperatures. Thus, elastic strain changes the universality class of the nematic phase transition (not of the magnetic transition!) but
does not destroy the nematic phase itself. 

The outline of the paper is as follows. In Section \ref{model} we
briefly introduce the model for spin-driven nematicity. In Section
\ref{phitheory} we develop a $\varphi^{4}$ theory for the nematic
fluctuations and estimate the Ginzburg regime, showing that nematic
fluctuations are expected to be large in the absence of the coupling
to the lattice. In Section \ref{strain} we include the nemato-elastic
coupling and analyze the nature of the nematic transition using a renormalization
group approach. We show that the coupling to the lattice introduces
non-analytic directional dependent terms in the propagator for nematic-fluctuations.
This results in softening only along certain directions in the momentum space. As a consequence, the upper critical
dimension becomes lower  compared to the case without coupling to the strain. We find $d_{uc}=2$
and the nematic transition becomes mean field for $d>2$. Finally, In Section \ref{conclusion}
we summarize our results.

\section{The model}

\label{model}
In what follows, we will work with 1-Fe unit cell with lattice constant $a$.
We begin by considering a model that includes coupled magnetic
and elastic degrees of freedom characterized by the action
\begin{equation}
S=S_{\Delta}+S_{\mathrm{ph}}+S_{c},
\end{equation}
with partition function 
\begin{equation}
Z=\int\mathcal{D}\mathbf{u}{\cal D}\left ( {\bf \Delta}_{x} {\bf \Delta}_{y} \right ) e^{-S}.
\end{equation}
Here, $S_{\Delta}$ represents the the action of the magnetic degrees of freedom, $S_{\mathrm{ph}}$ the phononic actions, and $S_{c}$ the coupling between magnetic degrees of freedom and phonons.
For the magnetic degrees of freedom we consider the model of a spin-driven
nematic phase discussed in details in Refs. \onlinecite{Fernandes12,
Una}, where the following expression for the magnetic action has
been derived:
\begin{eqnarray}
S_{\Delta} & = & \int_{x}\left[r_{0}\left({\bf \Delta}_{x}^{2}+{\bf \Delta}_{y}^{2}\right)+\left(\nabla{\bf \Delta}_{x}\right)^{2}+\left(\nabla{\bf \Delta}_{y}\right)^{2}\right]\nonumber \\
 & + & \int_{x}\left[\frac{u}{2N}\left({\bf \Delta}_{x}^{2}+{\bf \Delta}_{y}^{2}\right)^{2}-\frac{g}{2N}\left({\bf \Delta}_{x}^{2}-{\bf \Delta}_{y}^{2}\right)^{2}\right].\label{mag_action}
\end{eqnarray}
Here, $\int_{x}\cdots=\int d^{d}x\cdots.$ Similarly, we will use
below $\int_{q}\cdots=\int\frac{d^{d}q}{\left(2\pi\right)^{d}}\cdots$
for integrations over momenta.   ${\bf \Delta}_{x,y}$ denote
spin fluctuations associated with ${\bf Q}_{x,y}=\pi{\bf e}_{x,y}/a$
ordering wave-vectors respectively, see Fig 1. The most efficient approach to spin-driven nematicity is the large $N$ expansion, where $N$ denotes the number of components of the vectors   ${\bf \Delta}_{x,y}$ . The parameter $r_{0}$
tunes the distance from the magnetic phase transition, and $u$ and
$g$ denote the magnetic and nematic coupling constants respectively.
In a model based on itinerant electrons these constants can be expressed
as an integral over certain combinations of Green's functions across
different bands. \cite{Fernandes12, Una} Here we do not distinguish
between localized or itinerant magnetism and merely use the above
phenomenological form for the magnetic energy that is basically dictated
by symmetry as long as $g>0$. In our analysis we are using a continuum's
model. For an anisotropic system of moderately or weakly coupled layers
it is more appropriate to avoid the continuum's limit in the third
dimension. In Ref. \onlinecite{Fernandes12} such a model was analyzed
where $q^{2}\rightarrow q_{x}^{2}+q_{y}^{2}+A_{z}\left(1-\cos q_{z}\right)$,
where $q_{x,y}$ are still in the continuum's limit while $q_{z}$
goes from $-\pi$ to $\pi$. It was then shown that such an anisotropic
three dimensional system behaves very similar to a model in the continuum's
limit, yet with dimension $d$ intermediate between two and three,
i.e. $2<d<3$. In what follows we will pursue this continuum's approach
with variable dimensionality. 

The static phonon part of the action for the acoustic modes in the
$B_{1g}$-channel is given as 
\begin{equation}
S_{\mathrm{ph}}=N\int_{q}c_{s}(\mathbf{q})\left(q_{x}u_{x}-q_{y}u_{y}\right)^{2}.
\end{equation}
Here ${\bf u}=(u_x,u_y)$ is the phonon displacement field, $c_{s}({\mathbf{q}})$ is the sound velocity and ${\mathbf{q}}$ momentum along soft direction. $c_{s}({\mathbf{q}})$
is determined by the elastic constants of the material. It holds for a tetragonal
symmetry 
\begin{eqnarray}
c_{s}(\mathbf{q}) & = & c_{s}^{0}+\mu_{1} \cos^2{\theta}+\mu_{2}\sin^4{\theta}\sin^2{\left (2 \phi \right )},
\label{eq:sound vel}
\end{eqnarray}
where $c_{s}^{0}=c_{11}-c_{12}$ is the bare value of the sound velocity, and $\mu_{1}$,
and $\mu_{2}$ are expressed
in terms of the combinations of elastic constants of the system (for details
see the Appendix \ref{appb}). The coefficient $N$ in $S_{{\rm ph}}$
was introduced to generate a consistent expansion in large $N$.

Finally, the key magneto elastic coupling is 
\begin{equation}
S_{c}=\lambda_{\textrm{el}}\int_{x}\left({\bf \Delta}_{x}^{2}-{\bf \Delta}_{y}^{2}\right)\left(\partial_{x}u_{x}-\partial_{y}u_{y}\right),
\end{equation}
where $\lambda_{\textrm{el}}$ is the magneto-elastic coupling constant, and $\partial_{x}u_{x}-\partial_{y}u_{y}$ the orthorhombic distortion.

\begin{figure}
\begin{centering}
\includegraphics[width=1\columnwidth]{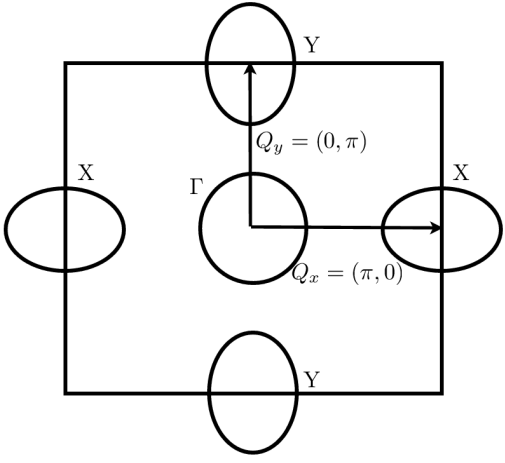} 
\par\end{centering}

\protect\caption{Band structure: the model consists of the central hole-like $\Gamma$
band, and the electron-like $X$ and $Y$ bands, shifted by ${\mathbf{Q}}_{X}=(\pi/a,0)$
and ${\mathbf{Q}}_{Y}=(0,\pi/a)$, respectively.}

\label{bandstructure} 
\end{figure}

\section{Collective nematic fluctuations and $\varphi^{4}$ theory of nematic
degrees of freedom}

\label{phitheory}

In order to develop a theory for collective nematic degrees of freedom,
we first perform a Hubbard-Stratonovich transformation of the two
quartic terms in the magnetic part $S_{\Delta}$ of the action Eq. (\ref{mag_action}) to obtain
\begin{eqnarray}
S_{\Delta}[\varphi, \lambda] & = & \frac{1}{2}\int_{x}\left(\frac{N}{u}\lambda^{2}+\frac{N}{g}\varphi^{2}\right)\nonumber \\
 & + & \int_{x}\left(\begin{array}{c}
{\bf \Delta}_{x}\\
{\bf \Delta}_{y}
\end{array}\right)^{T}{\cal G}^{-1}[\lambda,\varphi]\left(\begin{array}{c}
{\bf \Delta}_{x}\\
{\bf \Delta}_{y}
\end{array}\right),
\end{eqnarray}
where 
\begin{equation}
{\cal G}^{-1}[\lambda,\varphi]=\left(\begin{array}{cc}
r_{0}+i\lambda+\varphi-\nabla^{2} & 0\\
0 & r_{0}+i\lambda-\varphi-\nabla^{2}
\end{array}\right).
\end{equation}
We shift $\varphi\rightarrow\varphi+\lambda_{\textrm{el}}\left(\partial_{x}u_{x}-\partial_{y}u_{y}\right)$
and integrate out the magnetic modes. It follows 
\begin{eqnarray}
S & = & \frac{N}{2}\int_{x}\left(\frac{1}{u}\lambda^{2}+\frac{1}{g}\varphi^{2}\right)+N\int_{q}c_{s}(\mathbf{q})\left(q_{x}u_{x}-q_{y}u_{y}\right)^{2}\nonumber \\
 & - & \frac{\lambda_{\textrm{el}} N}{g}\int_{x}\varphi\left(\partial_{x}u_{x}-\partial_{y}u_{y}\right)+\frac{N}{2}{\rm tr}\log{\cal G}^{-1}\left[\lambda,\varphi\right],
\end{eqnarray}
where $c_{s}\left(\mathbf{q}\right)$ is given by Eq. (\ref{eq:sound vel})
with $c_{s} \rightarrow c_{s}+\lambda_{\textrm{el}}^{2}/g$. Finally, we
integrate over the phonon degrees of freedom, which leads to 
\begin{eqnarray}
S & = & \frac{N}{2}\int_{q}\left(\frac{1}{u}\lambda_{q}\lambda_{-q}+\left(\frac{1}{g}+\frac{\lambda_{\textrm{el}}^{2}}{c_{s}\left(q\right)}\right)\varphi_{q}\varphi_{-q}\right)\nonumber \\
 & + & \frac{N}{2}{\rm tr}\log{\cal G}^{-1}\left[\lambda,\varphi\right].
\end{eqnarray}
This action is an exact reformulation of our initial model. It is
the starting point of our subsequent analysis.

\begin{figure}
\begin{centering}
\includegraphics[width=1\columnwidth]{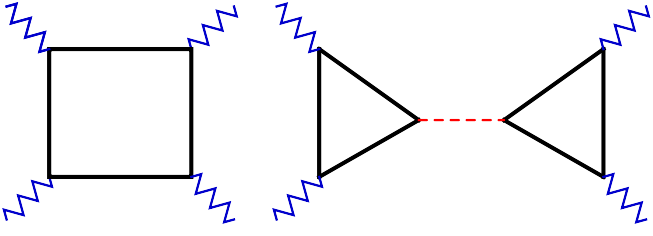} 
\par\end{centering}

\protect\caption{Diagrammatic contributions to $u_{\varphi}$. Full lines denote the
spin-fluctuation propagators $G_{0}$, wavy lines the propagators $D_{\varphi}$
and dotted lines the propagator $D_{\eta}$ (see the main text). The diagram on the left
is proportional to $L_{4}$ and gives a negative contribution to $u_{\varphi}$.
The diagram on the right contains two triangles $T$ given by Eq.
(\ref{T}), connected by a propagator $D_{\eta}$. This diagram becomes larger as the dimensionality
$d$ is lowered and it provides a positive contribution to $u_{\varphi}$.}

\label{triangles} 
\end{figure}

We concentrate on finite temperature transitions  and, for the moment focus on the tetragonal phase, where the nematic order
parameter is zero. We write 
\begin{eqnarray}
\lambda\left(x\right) & = & -i\psi_{0}+\eta\left(x\right),
\end{eqnarray}
where $\psi_{0}$ is determined by the saddle point equation that
becomes exact at leading order in $1/N$. Non-zero $\psi_{0}$ amounts
to a fluctuation renormalization of the magnetic correlation length.
$\eta\left(x\right)$ denotes the fluctuating part of $\lambda\left(x\right)$.
Similarly, for the Green's function matrix we can write 
\begin{equation}
{\cal G}^{-1}\left[\lambda,\varphi\right]={\cal G}_{0}^{-1}-{\cal V}\left[\eta,\varphi\right],
\end{equation}
where 
\begin{equation}
{\cal G}_{0}^{-1}=\left(r-\nabla^{2}\right)I,
\end{equation}
with $r=r_{0}+\psi_{0}$, and 
\begin{equation}
{\cal V}\left[\eta,\varphi\right]=-\left(\begin{array}{cc}
i\eta+\varphi & 0\\
0 & i\eta-\varphi
\end{array}\right).\label{nu}
\end{equation}
We expand:
\begin{eqnarray}
{\rm tr}\log{\cal G}^{-1}\left[\lambda,\varphi\right] & = & {\rm tr}\log{\cal G}_{0}^{-1}+{\rm tr}\log\left(1-{\cal G}_{0}{\cal V}\right)\nonumber \\
 & \approx & {\rm tr}\log{\cal G}_{0}^{-1}-\frac{1}{2}{\rm tr}\left({\cal G}_{0}{\cal V}\right)^{2}\nonumber \\
 &  & -\frac{1}{3}{\rm tr}\left({\cal G}_{0}{\cal V}\right)^{3}-\frac{1}{4}{\rm tr}\left({\cal G}_{0}{\cal V}\right)^{4}.
\end{eqnarray}
Now one can write the action as $S=S_0+S_2+S_3+S_4$, with
\begin{equation}
S_0= -\frac{N}{2}\int_{x}\frac{\psi_{0}^{2}}{u}+\frac{N}{2}{\rm tr}\log{\cal G}_{0}^{-1},
\end{equation}
and the quadratic part 
\begin{eqnarray}
S_{\mathrm{2}} & = & \frac{N}{2}\int_{q}\left(\begin{array}{c}
\eta\left(q\right)\\
\varphi\left(q\right)
\end{array}\right)^{T}D^{-1}\left(q\right)\left(\begin{array}{c}
\eta\left(-q\right)\\
\varphi\left(-q\right)
\end{array}\right),
\end{eqnarray}
where 
\begin{equation}
D^{-1}\left(q\right)=\left(\begin{array}{cc}
D^{-1}_{\eta}\left(q\right) & 0\\
0 & D^{-1}_{\varphi}\left(q\right)
\end{array}\right),
\end{equation}
with $D^{-1}_{\eta}\left(q\right)=\frac{1}{u}+\Pi\left(q, r\right)$ and 
$D^{-1}_{\varphi}\left(q\right)=\frac{1}{g}+\frac{\lambda^{2}}{c_{s}\left(q\right)}-\Pi\left(q, r\right)$.
The self-energy part is given by 
\begin{eqnarray}
\Pi\left(q,r\right) & = & \int_{k}\frac{1}{r+k^{2}}\frac{1}{r+\left({\bf k} +{\bf q} \right)^{2}}\nonumber \\
 & \approx & L_{2}r^{\frac{d}{2}-2}-b_{\varphi}q^{2},
\end{eqnarray}
with 
\begin{equation}
b_{\varphi}  =  -r^{\frac{d}{2}-3}\left(\frac{4}{d}L_{4}+\frac{d-4}{d}L_{3}\right),
\end{equation}
and we defined 
\begin{equation}
L_{n} = \int_{p}\frac{1}{\left(1+p^{2}\right)^{n}}=\frac{\Gamma\left(n-\frac{d}{2}\right)}{\left(2\sqrt{\pi}\right)^{d}\Gamma\left(n\right)}.
\end{equation}
The cubic action is given by
\begin{eqnarray}
S_{\mathrm{3}}=-N\int_{p,q}T(p,q)\varphi(-p)\varphi(p-q)\eta(q)
\end{eqnarray}
where 
\begin{eqnarray}
T(p,q) & = & i\int_{k}G_{0}(k)G_{0}(k-q)G_{0}(p+k-q)
\nonumber \\
\label{T}
\end{eqnarray}
is the triangular loop that appears in Fig. \ref{triangles}.

The quartic terms give the
following contribution to the action %
\begin{eqnarray}
S_{4} & = & -\frac{N}{4}L_{4}r^{d/2-4}\int_{q_1,q_2,q_3}\varphi_{q_1}\varphi_{q_2}\varphi_{q_3}\varphi_{-q_1-q_2-q_3}. \nonumber \\
\end{eqnarray}
Since the action $S$ is quadratic in $\eta$,
we can integrate the $\eta$-fields from the action, and use that
\begin{eqnarray}
 &  & \int\mathcal{D}\eta\exp\left\{ \int_{q}\left[-\frac{N}{2}\eta(q)D_{\eta}^{-1}(q)\eta(-q)+N j(q)\eta(q)\right]\right\} \nonumber \\
 & = & \exp{\left\{ -\frac{N}{2}j(q)D_{\eta}(q)j(-q)\right\} },
\end{eqnarray}
with $j(q)=\int_{p}T(p,q)\varphi(-p)\varphi(p-q)$.

The resulting field theory for the nematic fluctuations is therefore
of the form 
\begin{equation}
S_{\varphi}/N=\frac{1}{2}\int_{q}\left(r_{\varphi}+\frac{\lambda_{\rm el}^{2}}{c_{s}\left(\mathbf{q}\right)}+b_{\varphi}q^{2}\right)\varphi_{q}\varphi_{-q}+\frac{u_{\varphi}}{4}\int_{x}\varphi^{4},
\label{sphi}
\end{equation}
where 
\begin{eqnarray}
u_{\varphi} & = & -L_{4}r^{\frac{d}{2}-4}+\frac{2uL_{3}^{2}r^{d-6}}{1+uL_{2}r^{\frac{d}{2}-2}},\nonumber \\
r_{\varphi} & = & \frac{1}{g}-L_{2}r^{\frac{d}{2}-2},\nonumber \\
b_{\varphi} & = & r^{\frac{d}{2}-3}\left(\frac{4}{d}L_{4}+\frac{d-4}{d}L_{3}\right).\label{uphi}
\end{eqnarray}
We note that there
are two contributions to the quartic term $u_{\varphi}$; the mean-field
like term $\propto L_{4}$ which is negative, and the second term
$\propto L_{3}^{2}$, which is depicted in Figure \ref{triangles} on the right.
As we have shown above, the second term arises from the contraction
of $\eta$ propagators coming from two triangular loops $T$. This
term yields positive contribution and it is responsible for the change
in sign of the quartic coupling constant and thus for the possibility
to have second order split magnetic and nematic transitions. Analysing
both contributions shows that the term proportional to $L_3^2$ becomes increasingly
more important for lower space dimensionality $d$. Thus, low-dimensional
fluctuations are responsible for pre-emtive nematic order. If $d<4$,
$r^{\frac{d}{2}-2}\gg1$ as $r\rightarrow0$, and we get for
sufficiently large magnetic correlation length that 
\begin{eqnarray}
u_{\varphi} & = & \left(\frac{2L_{3}^{3}}{L_{2}}-L_{4}\right)r^{\frac{d}{2}-4}.
 \label{uphi2}
\end{eqnarray}
To avoid confusion, we stress that $r\propto \xi^{-2}$ determines the magnetic correlation length $\xi$, while $r_\varphi \propto \xi_\varphi^{-2}$ yields the nematic corrrelation length. The former vanishes at the critical point $T_N$ of  striped magnetic order, while the latter is zero at a nematic second order transition. We find that the quartic coupling constant $u_{\varphi}$, given by Eq. (\ref{uphi2}),
can therefore only be positive for $d<3$. Thus, only for very large
correlation length and $d<3$, is it possible that we obtain a second
order transition.  We remind the reader, that an anisotropic three dimensional system is described in terms of an intermediate dimensionality $2<d<3$, see Ref. \onlinecite{Fernandes12}.

Finally, we comment on the splitting between the nematic and the magnetic transitions. The condition for the second order nematic transition (for which $u_{\varphi}>0$ is required) to occur is that $\tilde r_{\varphi}=r_{\varphi}+\frac{\lambda^2_{\mathrm{el}}}{c_s^0} =0$, with $r_{\varphi}$ given by Eq. (\ref{uphi}). This gives
\begin{eqnarray}
\frac{1}{g}-L_2 r^{\frac{d}{2}-2}+\frac{\lambda^2_{\mathrm{el}}}{c_s^0}=0,
\end{eqnarray}
which occurs at the finite (but large) value of the magnetic correlation length $\xi \propto r^{-\frac{1}{2}}$. In other words, the nematic transition pre-empts the magnetic transition and occurs at a slightly higher temperature $T_s>T_N$. The temperature difference is dictated by the value of the nematic coupling constant $g$ and the size of the nemato-elastic coupling constant.

\subsection{The case without coupling to strain}

Let us first analyze the case $\lambda_{\textrm{el}}=0$ without coupling to the
lattice. We still need to determine whether, for a given set of coupling
constants $u,g$, the correlation length ever becomes large enough
to yield a positive sign for $u_{\varphi}$. In order to determine
the location of the tricritical point of the nematic degrees of freedom
(i.e. where $u_{\varphi}$ changes its sign), we use the fact that
at the nematic transition and for $\lambda_{\textrm{el}}=0$ holds that $r_\varphi=0$, i.e. $L_{2}r^{\frac{d}{2}-2}=\frac{1}{g}$,
which we solve to determine $r\left(g\right)$: 
\begin{equation}
r\left(g\right)=\left(\frac{2^{d}\pi^{d/2}}{\Gamma\left(2-\frac{d}{2}\right)g}\right)^{\frac{2}{d-4}}.
\end{equation}
We use this result to express $r$ in Eq. (\ref{uphi}) in terms of $g$
and obtain with $\alpha=u/g$: 
\begin{equation}
u_{\varphi}=\frac{4-d}{24 g^{\frac{8-d}{4-d}}}\frac{2\alpha\left(3-d\right)-\left(6-d\right)}{1+\alpha}\left(\frac{2^{d}\pi^{d/2}}{\Gamma\left(2-\frac{d}{2}\right)}\right)^{\frac{4}{4-d}}.
\end{equation}
Thus, if $d<3$ one only obtains a second order transition for 
\begin{equation}
\alpha>\alpha_{c}=\frac{6-d}{2\left(3-d\right)}.
\end{equation}
This result was obtained in Ref. \onlinecite{Fernandes12}
from an analysis of the equation of state of the nematic
order parameter.

Let us next estimate the size of the Ginzburg regime, i.e. the regime
of strong critical fluctuations of the nematic order parameter without
coupling to strain. One expects such critical nematic fluctuations
for $d\leq4$, if $u_{\varphi}>0$. The Ginzburg regime is most easily
estimated if we determine the natural dimensionless coupling constant
$\hat{u}_{\varphi}$. To this end we substitude $\varphi=\mu\phi$
and $q=\gamma k$, where $\gamma^{2}=r_{\varphi}/b_{\varphi}$ and
$\mu=b_{\varphi}^{\frac{d}{4}}r_{\varphi}^{-\frac{d+2}{4}}$ such
that 
\begin{equation}
S=\frac{1}{2}\int_{k}\left(1+k^{2}\right)\phi_{k} \phi_{-k}+\frac{\hat{u}}{4}\int_{k_{1,2,3}}\phi_{k_{1}}\phi_{k_{2}}\phi_{k_{3}}\phi_{-k_{1}-k_{2}-k_{3}},
\end{equation}
with dimensionless coupling constant: 
\begin{eqnarray}
\hat{u} & = & u_{\varphi}\gamma^{3d}\mu^{4}\nonumber \\
 & = & \frac{u_{\varphi}r_{\varphi}^{\frac{d-4}{2}}}{b_{\varphi}^{d/2}}.
\end{eqnarray}
We obtain 
\begin{eqnarray}
\hat{u}_{\varphi} & =\Upsilon_{d} & (gr_{\varphi})^{\frac{d-4}{2}},
\end{eqnarray}
with coefficient 
\begin{eqnarray}
\Upsilon_{d}= &  & \frac{\left ( 48 \pi \right )^{d/2}  \left (4-d \right)^{1-\frac{d}{2}}}{24 \Gamma \left ( 2-\frac{d}{2}\right )}
 \frac{2\alpha\left(3-d\right)-\left(6-d\right)}{\left(1+\alpha\right)}.\nonumber \\
\end{eqnarray}
Let us define $f=gr_{\varphi}$. If $f\rightarrow0$ the coupling
$\hat{u}_{\varphi}$ diverges. This is expected as we are below the
upper critical dimension. The Ginzburg regime $f_{Ginz}$ is determined
by $\hat{u}_{\varphi}(f)\approx1$. Thus, we analyze the coefficient
$\Upsilon_{d}\left(\alpha\right)$. Close to the tricritical point ($\alpha \approx \alpha_c$),
$\Upsilon_{d}\left(\alpha\right)$ vanishes like $\Upsilon_{d}\left(\alpha\right)=A(d) \frac{\alpha-\alpha_{c}}{\alpha_{c}}$, where $A(d)$ is the slope. 

\begin{figure}
\begin{centering}
\includegraphics[width=0.9\columnwidth]{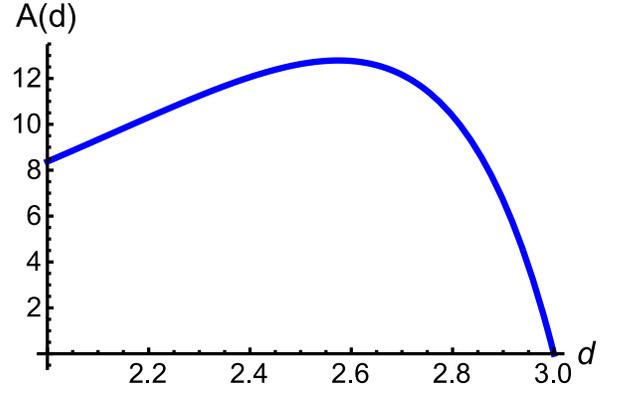} 
\par\end{centering}

\protect\caption{Slope $A(d)$ of the coefficient $\Upsilon_d$, near $\alpha_c$, for $2<d<3$. The dimensionality $2<d<3$ effectively describes $3$-dimensional, but anisotropic system, see Ref. \onlinecite{Fernandes12}. The slope is always bigger than $1$, except in the region around $d=3$. This signifies strong fluctuation regime (except near $d=3$).}

\label{slope} 
\end{figure}
Except for
$d$ near $d=3$, we find that the slope is always bigger than unity, see Fig. \ref{slope}. In this context
it is important to remind, that it was shown in Ref. \onlinecite{Fernandes12}
that for the phase diagram a dimensionality below $d$ effectively
describes a three-dimensional, yet anisotropic system. A slope bigger than unity already suggests a broad fluctuation regime, except for
a region near the tricritical point.

\subsection{Critical behavior with coupling to strain}

\label{strain}
Next, we include the coupling to strain, i.e. we analyze the action $S_{\varphi}$ of Eq. (\ref{sphi}) for finite $\lambda_{\rm el}$. This problem is similar to the one of a magnetic system with additional dipolar interactions. 
In Refs. \onlinecite{Aharony73, Aharony73B} the critical behaviour
of uniaxial magnets with dipolar interactions in $d$ dimensions was investigated
using renormalization group approach. It was found that in the presence
of the dipolar interaction the upper critical dimension is lowered from
$d=4$ (for the model including only exchange interaction) to $d=3$
(for the model containing both the exchange interaction and the dipolar
interaction). This is because the dipolar interactions generate non-analytic
directional-dependent terms in the spin propagator, similar to the non-analyticities that occur in the propagator in the case of second order elastic phase transitions, where the order parameter
is a component of the strain tensor and where an acoustic phonon emerges
as a soft mode.\cite{Folk76} Due to the directional dependence of
the propagator in elastic phase transition,\cite{Folk76} the softening occurs only along certain
directions, which leads to the mean field behaviour already in $d=3$
for systems where the softening occurs in $m=1$-dimensional subspace,
and for $d=3$ and $m=2$ logarithmic corrections were found to occur.

As seen from Eqs. (\ref{eq:sound vel}) and (\ref{sphi}), the coupling of the nematic fluctuations
to orthorhombic distortion, generates similar direction-dependent non-analytic terms in
the propagator for nematic fluctuations, and only certain directions in momentum space remain soft. 
In fact, for a tetragonal system, one finds that (see the Appendix \ref{appb} for the detailed derivation) the softening occurs in $m=1$-dimensional sub-manifold, in particular along the directions $q_x= \pm q_y$, $q_z=0$.

If we rescale the field to eliminate the coefficient $b_{\varphi}$, the action 
Eq. (\ref{sphi}) can be written in the following form
\begin{eqnarray}
S_{\varphi} =&&\frac{1}{2}\int_{q}\left(r_{\varphi}+\frac{\lambda^2_{\mathrm{el}}}{c_s(q)}+q^{2} \right)
\varphi_q \varphi_{-q} \nonumber \\
&+& \frac{u_{\varphi}}{4}\int_{q_1,q_2,q_3}\varphi_{q_1}\varphi_{q_2}\varphi_{q_3}\varphi_{-q_1-q_2-q_3},
\end{eqnarray}
with $c_s(q)$ given by Eq. (\ref{eq:sound vel}) and $\mu_1,\mu_2$ are given in Appendix \ref{appb}.
We analyze this action using one a loop renormalization group approach. We define $\tilde r_{\varphi}=r_{\varphi}+\frac{\lambda_{\textrm{el}}^2}{c_s^{0}}$. The flow equations for this $\varphi^{4}$ theory are straightforward, and we obtain the usual result (see for example Ref. \onlinecite{Chaikin}):
\begin{eqnarray}
\frac{d\tilde{r}_{\varphi}}{dl} & = & 2\tilde{r}_{\varphi}+3u_{\varphi}\frac{d}{dl}\int_{q}^{>}D_{\varphi}\left(q\right),\nonumber \\
\frac{du_{\varphi}}{dl} & = & \left(4-d\right)u_{\varphi}-9u_{\varphi}^{2}\frac{d}{dl}\int_{q}^{>}D^2_{\varphi}\left(q\right).
\end{eqnarray}
The momentum integration is performed over momenta between $\Lambda/b<q<\Lambda$,
where $\Lambda$ is the cut-off and $b=e^{-l}$. The propagator for
nematic fluctuations in the presence of the coupling to the lattice
is given by 
\begin{eqnarray}
D^{-1} _{\varphi}\left(q\right)& \approx & \tilde{r}_{\varphi}+q^{2}+
h_2\Lambda^{2} \sin^4{\theta}\sin^2{2 \phi} +h_1\Lambda^{2}  \cos^2{\theta}, \nonumber \\
\end{eqnarray}
where we introduced $h_{i}\Lambda^{2}=-\frac{\lambda^{2}}{\left(c_{s}^{0}\right)^{2}}\mu_{i}.$
Simple power counting arguments show that the coupling constants $h_{i}$
are relevant and they grow according to 
\begin{equation}
\frac{dh_{i}}{dl}=2h_{i}.
\end{equation}
This flow equation will not be modified by interaction corrections,
as the elimination of high energy modes cannot generate non-analytic
corrections of the type $q_{z}^{2}/q^{2} \sim  \cos^2{\theta}$ or $q_{x}^{2}q_{y}^{2}/q^{4} \sim \sin^4{\theta}\sin^2{2 \phi}$. Thus, we have 
\begin{equation}
h_{i}\left(l\right)=h_{i}e^{2l}.
\end{equation}
In addition we have 
\begin{eqnarray}
\frac{d\tilde{r}_{\varphi}}{dl} & = & 2\tilde{r}_{\varphi}+3u_{\varphi}\Lambda^{d-2}A\left(h_{1},h_{2}\right)-3u_{\varphi}r\Lambda^{d-4}B\left(h_{1},h_{2}\right),\nonumber \\
\frac{du_{\varphi}}{dl} & = & \left ( 4- d \right ) u_{\varphi}-9u_{\varphi}^{2}\Lambda^{d-4}B\left(h_{1},h_{2}\right),
\end{eqnarray}
with 
\begin{eqnarray}
A\left(h_{1},h_{2}\right) & = & K_{d-1}\int \frac{\sin^{d-2}{\theta}d\theta d\phi}{\left(2\pi\right)^{2}}f(\theta,\phi),\nonumber \\
B\left(h_{1},h_{2}\right) & = & K_{d-1}\int \frac{\sin^{d-2}{\theta}d\theta d\phi}{\left(2\pi\right)^{2}}f^{2}(\theta,\phi).\label{AB}
\end{eqnarray}
We used $\int \cdots=\int_{0}^{\pi}d\theta\int_{0}^{2\pi}d\phi\cdots$
as well as
\begin{equation}
f(\theta,\phi)=\frac{1}{1+h_{1}\cos^{2}\theta+h_{2}\sin^{4}\theta\sin^{2}\left(2\phi\right)}.
\end{equation}
Further more $K_{d}=\frac{2\pi^{d/2}}{\left(2\pi\right)^{d}}/\Gamma\left(\frac{d}{2}\right)$.
For large $h_{i}$ the main contribution to the integrals in Eq.(\ref{AB})
comes from the vicinity $\theta\approx\pi/2$ and $\phi\approx0$
and one finds that 
\begin{eqnarray*}
A\left(h_{1},h_{2}\right) & \sim & \frac{1}{2}\frac{K_{d-1}}{(2 \pi)^2}\left(h_{1}h_{2}\right)^{-1/2},\\
B\left(h_{1},h_{2}\right) & \sim &\frac{1}{4}  \frac{K_{d-1}}{(2 \pi)^2}\left(h_{1}h_{2}\right)^{-1/2}.
\end{eqnarray*}
Therefore we introduce
the effective coupling constant 
\begin{equation}
G=\frac{u_{\varphi}}{\sqrt{h_{1}h_{2}}},
\end{equation}
such that 
\begin{equation}
\frac{dG}{dl}=\left(2-d\right)G-9G^{2}\Lambda^{d-4} \frac{K_{d-1}}{4(2 \pi)^2}.
\end{equation}
which flows to zero for $d>2$, leading to mean field behaviour above $d=2$.\cite{Levanyuk70,Cowley76,Folk76}

Next, we would like to estimate the temperature range in which the mean-field behaviour can be expected, and in which the Currie-Weiss behavior of nematic degrees of freedom can be observed. The scaling $A,B \simeq (h_1 h_2)^{-1/2}$ breaks down when one of the $h_i$ becomes of order $1$, see Eq. (\ref{AB}). Let us assume that this happens at length $l=l^*$, such that $h_1(l^*) \approx 1$. One finds that $h_2(l^*)=\frac{h_2(0)}{h_1(0)}$, and that for $d=2$,  $u_{\varphi}(l^*)=\frac{u_{\varphi}(0)}{h_1(0)}$. The nematic correlation length is given by $\xi_{\varphi}(l^*)=\sqrt{h_1(0)} \xi_{\varphi}(0)$. The analysis will break, down when the correlation length becomes smaller than the lattice spacing $a$, i.e. for $\xi_{\varphi}(l^*) < a$. Therefore, we expect the mean-field behavior to be valid only for
\begin{eqnarray}
\xi_{\varphi}(0) > \frac{a \Lambda c_s^{0}}{\sqrt{\lambda_{\mathrm{el}}^2 \mu_1}} \approx \sqrt{\frac{c_s^{0}}{\lambda_{\mathrm{el}}^2}},
\label{xiestimate}
\end{eqnarray}
where $\lambda_{\mathrm{el}}$ is the nemato-elastic coupling constant, and $c_s^0$ and $\mu_1$ represent the combinations of various elastic constants of the material (see Appendix A). In Eq. (\ref{xiestimate}) we have used that $\mu_i \sim c_s^{0}$, see Ref. \onlinecite{elastic} for experimental data, and that $a \Lambda \sim 1$.
The susceptibility of the nematic order parameter is given by $\chi_{\varphi}=r_{\varphi}^{-1}= \xi_{\varphi}^2$. Therefore, the condition for the mean-field behavior reduces to the following condition for the nematic susceptibility 
\begin{eqnarray}
\chi_{\varphi} > \frac{c_s^0}{\lambda_{\mathrm{el}}^2}.
\label{cw}
\end{eqnarray}
In Ref. \onlinecite{Fernandes2010}, it was shown that the elastic modulus $c_s$ softens as one approaches the structural transition and that it effectively measures the nematic susceptibility through
\begin{eqnarray}
c_s^{-1}=\left ( c_s^{0} \right)^{-1} \left ( 1 + \frac{\lambda_{\mathrm{el}}^2}{c_s^{0}}\chi_{\varphi} \right ).
\label{modulus}
\end{eqnarray}
We see that when the nematic susceptibility becomes of the order $\chi_{\varphi} \sim \frac{c_s^0}{\lambda_{\mathrm{el}}^2}$ strong renormalisation of the elastic modulus will take place. Therefore, we have explicitly shown that the mean field behavior, Eq. (\ref{cw}) occurs in the entire regime where the supression of the elastic modulus takes place, see Fig. \ref{currie} for details. This is exactly what has been observed experimentally. \cite{Anna1, Anna3, yoshizawa} We add that this conclusion is valid regardless of the detailed microscopic origin of nematicity. Curie-Weiss behavior due to the coupling to the lattice is expected in the entire temperature regime where a softening of the elastic constant is observed. For several iron based systems, this regime can be as high as 300-350 K. \cite{Fernandes2010, Anna1, Anna3, yoshizawa}

\begin{figure}
\begin{centering}
\includegraphics[width=1\columnwidth]{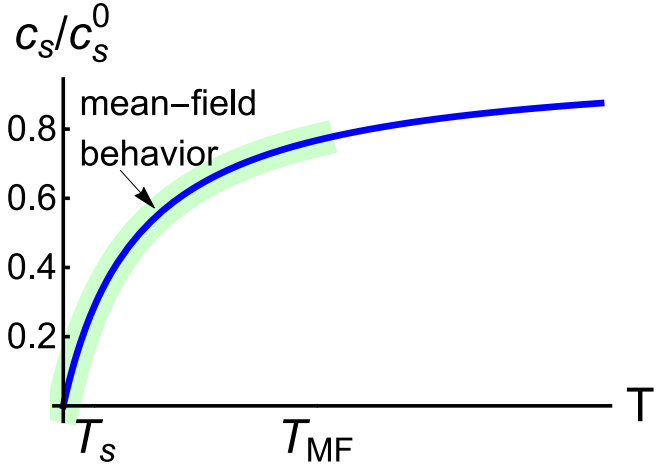} 
\label{currie}
\par\end{centering}

\protect\caption{The regime in which the mean-field behaviour of the nematic degrees of freedom can be expected (shaded light green region). The figure shows the softening of the elastic modulus $c_s$, Eq. (\ref{modulus}), denoted by a solid blue line, as one approaches the structural transition from the high-temperature tetragonal phase. $T_{\mathrm{MF}}$ indicates the characteristic temperature where mean field behavior of nematic degrees of freedom sets in upon approaching the transition at $T_s$. The mean-field regime coincides with the regime of strong elastic modulus renormalization (see the main text for the explanation).}

\label{currie} 
\end{figure}

\section{Conclusion}

\label{conclusion}
In the spin-driven nematic scenario magnetic fluctuations associated with the striped magnetic order cause the formation of the nematic state -- the state with no magnetic order, but broken $Z_2$ symmetry. Therefore, fluctuations are crucial for the existence of the nematic state. However, experimentally it has been measured that the nematic degrees of freedom behave mean-field like in a very broad temperature range --  in particular, the Currie-Weiss dependence of the nematic susceptibility was observed. On the first sight, these two observations might seem to be in contradiction. 

The present paper reconciles these two statements and determines the temperature regime where a Curie-Weiss behavior of the nematic susceptibility is expected. In particular, we show that the coupling to the lattice suppresses the fluctuations of the nematic order parameter itself, which renders the nematic transition mean-field, but it does not affect the magnetic transition or the very existence of the nematic phase. Starting from a microscopic model of a spin-driven nematic phase, which also explains the emergence of the nematic phase at a slightly higher temperature than the Neel temperature (separated magnetic and nematic transitions), we constructed the $\varphi^4$ theory of the nematic degrees of freedom. 
First we ignored the coupling to the lattice and by analyzing the 
quartic coefficient, we showed that nematic fluctuations are characterized by a rather large Gizburg regime. Next, we added the  coupling between nematic degrees of freedom to elastic strain to the $\varphi^4$ theory and analyzed this using the renormalization group procedure. We have found that, due to the nemato-elastic coupling which introduces directional-dependent terms in the propagator for nematic fluctuations rendering only certain directions to become soft, the nematic transition becomes mean-field for $d>2$. Most importantly, the nemato-elastic coupling does not supress fluctuations that cause the nematic order in the first place (i.e. magnetic fluctuations), it only supresses the fluctuations of the nematic order parameter itself. We have found that the nematic transition happens at large but finite magnetic correlation length, such that one obtains split magnetic and nematic transitions, with the nematic transition being mean-field like (rather than in the Ising universality class), while the magnetic transition is expected to behave in a non mean-field like fashion.

Finally, we found that the mean-field behaviour of the nematic degrees of freedom should occur in the entire regime where there is a significant softening of the elastic modulus. This is in excellent agreement with the experiments, where Currie-Weiss behaviour of the nematic susceptibility was measured across a rather large temperature range \cite{Fernandes2010, Anna1, Anna3, yoshizawa} (up to 300-350$K$ for some iron-based superconductors), which coincides with the temperature range in which a significant reduction of the elastic modulus was observed.

\section{Acknowledgement}

We acknowledge useful discussions with A. Chubukov, R. Fernandes, I. Paul and M. Sch\"utt.
U.K. acknowledges the support from the Helmholtz Association, through
Helmholtz post-doctoral grant PD-075 ``Unconventional order and superconductivity
in pnictides''. J.S. acknowledges the support from Deutsche Forschungsgemeinschaft
(DFG) through the Priority Program SPP 1458 ``Hochtemperatur-Supraleitung
in Eisenpniktiden'' (project-no. SCHM 1031/5-1).

\appendix

\begin{appendix}
\section{Momentum dependence of the elastic constant $c_s(q)$}
\label{appb}
The elastic part of the free energy of a tetragonal system is given by
\begin{eqnarray}
F_{\mathrm{el}} &=& \frac{c_{11}}{2} \left ( \epsilon_{xx}^2 + \epsilon_{yy}^2\right ) + 
\frac{c_{33}}{2}\epsilon_{zz}^2 
+\frac{c_{44}}{2} \left ( \epsilon_{xz}^2 + \epsilon_{yz}^2\right )
\nonumber \\
&+& \frac{c_{66}}{2}\epsilon_{xy}^2 
+c_{12} \epsilon_{xy} \epsilon_{yy}
+c_{13}   \left ( \epsilon_{xx} + \epsilon_{yy}\right ) \epsilon_{zz},
\label{elfree}
\end{eqnarray}
where $\epsilon_{ij}=\frac{\partial_{i}u_j+\partial_{j}u_i}{2}$, and $u_i$ is the $i$th component of the phonon displacement field. 

The dynamic matrix $M$ is defined from $F_{\mathrm{el}} = \frac{1}{2} u_i (q) M_{ij}(q) u_{j}(q)$. It can be expressed as
\begin{eqnarray}
M_{ij}=\sum_{m,l} q_{m} q_{l} \frac{ \partial^2 F_{\mathrm{el}} } {\partial \epsilon_{ik}\partial \epsilon_{lj}}.
\end{eqnarray}
For a tetragonal system the dynamic matrix is given by
\begin{widetext} 
\begin{eqnarray}
M ({\mathbf{q}})=\left(\begin{array}{ccc}
c_{11}q_{x}^{2}+c_{66}q_{y}^{2}+c_{44}q_{z}^{2} & \left(c_{12}+c_{66}\right)q_{x}q_{y} & \left(c_{13}+c_{44}\right)q_{x}q_{z}\\
\left(c_{12}+c_{66}\right)q_{x}q_{y} & c_{66}q_{x}^{2}+c_{11}q_{y}^{2}+c_{44}q_{z}^{2} & \left(c_{13}+c_{44}\right)q_{y}q_{z}\\
\left(c_{13}+c_{44}\right)q_{x}q_{z} & \left(c_{13}+c_{44}\right)q_{y}q_{z} & c_{44}\left(q_{x}^{2}+q_{y}^{2}\right)+c_{33}q_{z}^{2}
\end{array}\right).
\end{eqnarray}
\end{widetext} 
The phonon frequencies $\omega$ in a tetragonal system can be determined from the dynamic matrix $M$, via $\mathrm{det} \left ( \omega^2 \rho - M ({\mathbf{q}})\right )=0$, where $\rho$ denotes the density. 
A vanishing elastic constant corresponds to a vanishing sound velocity. Here, we are interested in the case $c_{11}-c_{12}\rightarrow 0$.

The soft directions, along which the sound velocity vanishes correspond to the two lines in the $xy$ plane (i.e. $q_z=0$): $q_x=q_y$, and $q_x=-q_y$. Along these directions we have that $\omega^2 \rho = (c_{11}-c_{12})q_x^2 \rightarrow 0$. Now, if one calculates the dispersion in the vicinity of such line, for example $q_x=q_y$ one finds that

\begin{eqnarray}
\omega^2 \rho & \approx & \left (c_{11}-c_{12} \right ) \frac{\left ( q_x+q_y \right )^2}{2} \nonumber \\ 
&+& 
\frac{\left (q_x-q_y \right )^2}{2} \left [c_{11} + c_{66} - \frac{\left ( c_{11}-c_{66}\right )^2}{c_{12}+c_{66}} \right ]
+c_{44} q_z^2, \nonumber \\
\end{eqnarray}
where $q_z \ll q_x $ and $q_x-q_y \ll q_x$. Choosing the angle parametrization such that $q_x=q \sin{\theta}\cos{\left ( \phi+ \frac{\pi}{4} \right )}$, and $q_y=q \sin{\theta}\sin{\left ( \phi+ \frac{\pi}{4} \right )}$, we get $\omega^2 \rho  \approx  c_s(\mathbf{q}) q^2_{+}$, 
with $q_+=(q_x+q_y)$ being the soft momentum and
\begin{eqnarray}
c_s(\mathbf{q}) = \frac{\left (c_{11}-c_{12} \right )}{2}+
\mu_1 \sin^4{\theta}\sin^2{2 \phi} +\mu_2 \cos^2{\theta},
\end{eqnarray}
with
\begin{eqnarray}
\mu_1 &=&\frac{1}{8} \left [c_{11} + c_{66} - \frac{\left ( c_{11}-c_{66}\right )^2}{c_{12}+c_{66}} \right ],
\nonumber \\
\mu_2 &=& c_{44}.
\end{eqnarray}

\end{appendix}


\begin{thebibliography}{0}
\expandafter\ifx\csname natexlab\endcsname\relax\def\natexlab#1{#1}\fi
\expandafter\ifx\csname bibnamefont\endcsname\relax
  \def\bibnamefont#1{#1}\fi
\expandafter\ifx\csname bibfnamefont\endcsname\relax
  \def\bibfnamefont#1{#1}\fi
\expandafter\ifx\csname citenamefont\endcsname\relax
  \def\citenamefont#1{#1}\fi
\expandafter\ifx\csname url\endcsname\relax
  \def\url#1{\texttt{#1}}\fi
\expandafter\ifx\csname urlprefix\endcsname\relax\def\urlprefix{URL }\fi
\providecommand{\bibinfo}[2]{#2}
\providecommand{\eprint}[2][]{\url{#2}}

\end{thebibliography}


\begin{thebibliography}{0}
\expandafter\ifx\csname natexlab\endcsname\relax\def\natexlab#1{#1}\fi
\expandafter\ifx\csname bibnamefont\endcsname\relax
  \def\bibnamefont#1{#1}\fi
\expandafter\ifx\csname bibfnamefont\endcsname\relax
  \def\bibfnamefont#1{#1}\fi
\expandafter\ifx\csname citenamefont\endcsname\relax
  \def\citenamefont#1{#1}\fi
\expandafter\ifx\csname url\endcsname\relax
  \def\url#1{\texttt{#1}}\fi
\expandafter\ifx\csname urlprefix\endcsname\relax\def\urlprefix{URL }\fi
\providecommand{\bibinfo}[2]{#2}
\providecommand{\eprint}[2][]{\url{#2}}



\bibitem{Birgeneau11}
 C.~R. Rotundu,  and
 R.~J. Birgeneau,
  Phys. Rev. B
  \textbf{84}, 092501 (2011).

\bibitem{Kim11}
 M.~G. Kim,
 R.~M. Fernandes,
 A. Kreyssig,
J.~W. Kim, 
A. Thaler,  
S.~L. Bud'ko,  
P.~C. Canfield, 
R.~J. McQueeney,  
J. Schmalian,    and
A.~I. Goldman, 
  Phys. Rev. B
  \textbf{83}, 134522 (2011).
  
  \bibitem{matsuda_t}
 S. Kasahara,
 H.~J. Shi,
 K. Hashimoto,
S. Tonegawa, 
Y. Mizukami,  
T. Shibauchi,  
K. Sugimoto, 
T. Fukuda,  
T. Terashima,   
A.~H. Nevidomskyy,  and
 Y. Matsuda, 
  Nature
  \textbf{486}, 382 (2012).
  
  \bibitem{nematic_review}
 R.~M. Fernandes,  and
 J. Schmalian,
 Supercond. Sci. Technol.
  \textbf{25},  084005 (2012).
  
  \bibitem{Chu10}
 J.~H. Chu,
 J.~G. Analytis,
 K. De Greve,
 P.~L. McMahon,
 Z. Islam,
   Y. Yamamoto,  and
   I.~R. Fisher,
 Science
  \textbf{329}, 824 (2010).

  
  \bibitem{Tanatar10}
 M.~A. Tanatar,
   E.~C. Blomberg,
 A. Kreyssig,
   M.~G. Kim,
   N. Ni,
   A. Thaler,
   S.~L.  Bud'ko,
       P.~C. Canfield,
       A.~I. Goldman,
       I.~I. Mazin,  and
         R. Prozorov,
  Phys.
Rev. B
  \textbf{81}, 184508 (2010).
  
  \bibitem{Chu2012}
 J.~H. Chu,
  H.~H. Kuo,
 J.~G. Analytis,  and
   I.~R. Fisher,
 Science
  \textbf{337}, 719 (2012).


  

  \bibitem{Fernandes13_shear}
 R.~M. Fernandes,
 A.~E. B\"ohmer,
 C. Meingast,  and
 J. Schmalian,
 Phys. Rev. Lett.
  \textbf{111}, 137001 (2013).
  
  \bibitem{Fernandes2010}
 R.~M. Fernandes,
   L.~H. VanBebber,
 S. Bhattacharya,
   P. Chandra,
   V. Keppens,
   D. Mandrus,
   M.~A. McGuire,
       B.~C. Sales,
       A.~S. Sefat,  and
       J. Schmalian,
  Phys.
Rev. Lett.
  \textbf{105}, 157003 (2010).

\bibitem{Kontani1}
 H. Kontani,  and
 Y. Yamakawa,
 Phys. Rev. Lett.
  \textbf{113}, 047001 (2014).

\bibitem{Kontani2}
 H. Kontani,
 T. Saito,  and
   S. Onari,
  Phys. Rev. B
  \textbf{84}, 024528 (2011).


\bibitem{Anna1}
 A.~E. B\"ohmer,
 P. Burger,
 F. Hardy,
 T. Wolf,
 P. Schweiss,
 R. Fromknecht,
 M. Reinecker,
 W. Schranz,  and
 C. Meingast,
 Phys. Rev. Lett.
  \textbf{112}, 047001 (2014).


\bibitem{Gallais}
 Y. Gallais,
   R.~M. Fernandes,
     I. Paul,
 L. Chauviere,
     Y.~X. Yang,
 M.~A. Measson,
 M. Cazayous,
   A. Sacuto,
       D. Colson,  and
       A. Forget,
 Phys. Rev. Lett.
 \textbf{111}, 267001
  (2013).

  
  \bibitem{Rudi}
 F. Kretzschmar,
   T. B\"ohm,
 U. Karahasanovic,
   B. Muschler,
   A. Baum,
   D. Jost,
     J. Schmalian,  
     S. Caprara,
     M. Grilli,
     C. Di Castro,
     J. Analytis,
     J. Chu,
      I.~R. Fisher,
      and
      R. Hackl,     
arXiv:1504.04570,
  (2015).
  
  \bibitem{Blumberg1}
 W.~L. Zhang,
 P. Richard,
 H. Ding,
 Athena S. Sefat,
   J. Gillett,
 Suchitra E. Sebastian,
 M. Khodas,  and
 G. Blumberg,
 arXiv:1410.6452, (2015).
  
  \bibitem{Blumberg2}
 V.~K. Thorsmolle,
   M. Khodas,
     Z.~P. Yin,
 Chenglin Zhang,
     S.~V. Carr,
 Pengcheng Dai,  and
 G. Blumberg,
  arXiv:1410.6456, (2015).
  
   \bibitem{Khodas15}
 M. Khodas,
  and
   A. Levchenko,
  Phys. Rev. B
  \textbf{91}, 235119 (2015).
  
  \bibitem{Una}
 U. Karahasanovic,
 F. Kretzschmar,
   T. B\"ohm,
 R. Hackl,
 I. Paul,
 Y. Gallais,
  and
   J. Schmalian,
  Phys. Rev. B
  \textbf{92}, 075134 (2015).
  
  
   \bibitem{yann_review}
   Y. Gallais,
     and
 I. Paul,
  arXiv:1508.01319
  (2015).
  
  
   \bibitem{Gallais15}
   Y. Gallais,
     I. Paul,
       L. Chauviere,
     and
 J. Schmalian,
 arXiv:1504.04570, 
  (2015).
  
  \bibitem{Jiang2013}
 Shuai Jiang,
 H.~S. Jeevan,
 Jinkui Dong,  and
 P. Gegenwart,
 Phys. Rev. Lett.
  \textbf{110}, 067001 (2013).

\bibitem{Dusza2011}
 A. Dusza,
 A. Lucarelli,
 F. Pfuner,
 J.~H. Chu,
 I.~R. Fisher,  and
 L. Degiorgi,
Europhys. Lett.
  \textbf{93}, 37002 (2011).

\bibitem{Nakajima2011}
 M. Nakajima,
 T. Liang,
 S. Ishida,
 Y. Tomioka,
   K. Kihou,
 C.~H. Lee,
 A. Iyo,
   H. Eisaki,
   T. Kakeshita,
   T. Ito,  and
   S. Uchida,
Proc. Natl. Acad. Sci. U.S.A.
  \textbf{108}, 12 238 (2011).

\bibitem{Rosenthal13}
 E.~P. Rosenthal,
 E.~F. Andrade,
 C.~J. Arguello,
 R.~M. Fernandes,
   L.~Y. Xing,
 X.~C. Wang,
 C.~Q. Jin,
   A.~J. Millis,  and
   A.~N. Pasupathy,
  Nature Phys.
  \textbf{10}, 225232 (2014).
  

  
  \bibitem{Phillips11}
 W. Lv,  and
   P. Phillips,
  Phys. Rev. B
 \textbf{84}, 174512
  (2011).
  
  \bibitem{Applegate11}
 R. Applegate,
   R.~R.~P. Singh,
     C.~C. Chen,  and
 T.~P. Devereaux,
  Phys. Rev. B
 \textbf{85}, 054411
  (2012).
  
  \bibitem{Dagotto13}
 S. Liang,
   A. Moreo,  and
   E. Dagotto,    
 Phys. Rev. Lett.
 \textbf{111}, 047004
  (2013).

\bibitem{w_ku10}
 C.~C Lee,
   W.~G. Yin,  and
   W. Ku,    
 Phys. Rev. Lett.
 \textbf{103}, 267001
  (2009).

  
  \bibitem{kruger1}
 F. Kr\"uger,
   S. Kumar,
   J. Zaanen,    and
       J. van den Brink,     
  Phys. Rev. B
 \textbf{79}, 054504
  (2009).
  
\bibitem{kruger2}
   W. Lv,
 F. Kr\"uger,  and
   P. Phillips,   
  Phys. Rev. B
 \textbf{82}, 045125
  (2010).
  
  
  \bibitem{Fernandes12}
 R.~M. Fernandes,
   A.~V. Chubukov,
 J. Knolle,
 I. Eremin,  and
 J. Schmalian,
  Phys. Rev. B
  \textbf{85}, 024534 (2012).
  
  
  \bibitem{naturereview}
 R.~M. Fernandes,
   A.~V. Chubukov,  and
 J. Schmalian,
  Nature Phys.
  \textbf{10}, 97-104 (2014).


\bibitem{Xu08}
 C. Xu,
   M. M\"uller,  and
 S. Sachdev,
  Phys. Rev. B
  \textbf{78}, 020501(R) (2008).

\bibitem{Fang08}
 C. Fang,
   H. Yao,
 W.~F. Tsai,
   J.~P. Hu,  and
S.~A. Kivelson,
  Phys. Rev. B
  \textbf{77}, 224509 (2008).

\bibitem{Qi09}
 Y. Qi,  and
   C. Xu,
  Phys. Rev. B
  \textbf{80}, 094402 (2009).

\bibitem{Cano10}
 A. Cano,
   M. Civelli,
 I. Eremin,  and
   I. Paul,
  Phys. Rev. B
  \textbf{82}, 020408(R) (2010).
  
  \bibitem{Terashima15}
 T. Terashima,
   N. Kikugawa,
 S. Kasahara,  
  T. Watashige, 
  T. Shibauchi, 
   Y. Matsuda, 
    A.~E. B\"ohmer, 
     F. Hardy, 
      C. Meingast,
        H.  v L\"ohneysen, 
  and
   S. Uji,
  J. Phys. Soc. Jpn.
  \textbf{84}, 063701 (2015).
  
   \bibitem{Anna15}
 A.~E. B\"ohmer,
   T. Arai,
 F. Hardy,  
  T. Hattori, 
  T. Iye, 
   T. Wolf, 
    H. v. L\"ohneysen, 
     K. Ishida, 
       and
      C. Meingast,
 Phys. Rev. Lett.
  \textbf{114}, 027001 (2015).
  
    \bibitem{Baek15}
 S.~H. Baek,
   D.~V. Efremov,
 J.~M. Ok,  
  J.~S. Kim, 
  T. Iye, 
   J. van den Brink, 
      and
    B. B\"uchner, 
 Nature Materials
  \textbf{14}, 210214 (2015).
  
   \bibitem{Watson15}
 M.~D. Watson,
   T.~K. Kim,
 A.~A. Haghighirad,  
  N.~R. Davies, 
  A. McCollam, 
   A. Narayanan, 
    S.~F. Blake, 
     Y.~L. Chen, 
      S. Ghannadzadeh, 
       A.~J. Schofield, 
       M. Hoesch, 
         C. Meingast, 
           T. Wolf,   
      and
    A.~I. Coldea, 
arXiv:1502.02917
 (2015).
 
 
  \bibitem{Terashima14}
 T. Terashima, 
      and
    et. al.,
 Phys. Rev. B
  \textbf{90}, 144517 (2014).


  \bibitem{Chubukov15}
 A.~ V. Chubukov,
   R.~ M. Fernandes,
  and
   J. Schmalian,
  Phys. Rev. B
  \textbf{91}, 201105 (2015).
  
   \bibitem{RG1}
 A.~ V. Chubukov,
   D. Efremov,
  and
   I. Eremin,
  Phys. Rev. B
  \textbf{78}, 134512 (2008).
  
 

 \bibitem{RG2}
 A.~ V. Chubukov,
   \bibinfo{journal}{Physica C}
  \textbf{469}, 640 (2009).
 
   \bibitem{RG3}
    S. Maiti,
      and
 A.~ V. Chubukov,
  Phys. Rev. B
  \textbf{82}, 214515 (2010).

  
  
    \bibitem{Anna3}
 A.~E. B\"ohmer,
   and
 C. Meingast,
arXiv:1505.05120  
  (2015).
  
    \bibitem{yoshizawa}
 M. Yoshizawa,
  et. al.
  J. Phys. Soc. Jpn
  \textbf{81}, 024604 (2012).
  
  
  \bibitem{Levanyuk70}
 A.~P. Levanyuk,
   and
   A.~A. Sobyanin,
 JETP
  \textbf{11}, 371 (1970).
  
 
   \bibitem{Cowley76}
 R.~A. Cowley,
  Phys. Rev. B
  \textbf{13}, 4887 (1976).
  
    \bibitem{Folk76}
 R. Folk,
 H. Iro,
   and
   F. Schwabl,
  Z. Phys. B 25
  \textbf{69}, 371 (1976).
  
  
  
   \bibitem{Folk79}
 R. Folk,
 H. Iro,
   and
   F. Schwabl,
  Phys. Rev. B
  \textbf{20}, 1229 (1979).
  


\bibitem{Zacharias15}
 M. Zacharias,
   I. Paul,
  and
   M. Garst,
 Phys. Rev. Lett.
  \textbf{115}, 025703 (2015).


   \bibitem{Aharony73}
 A. Aharony,
   and
 M. Fisher,
  Phys. Rev. B
  \textbf{8}, 3323 (1973).
  
   \bibitem{Aharony73B}
 A. Aharony,
  Phys. Rev. B
  \textbf{8}, 3363 (1973).
  
      
       \bibitem{Chalker80}
 J.~T. Chalker,
 Phys. Lett.
  \textbf{80A}, 4042 (1980).
  
  
     \bibitem{Chaikin}
 P.~M. Chaikin,
  and
   T.~C. Lubensky,
  Principles of condensed matter physics,
  Cambridge University Press (October 9, 2000).
   
 
    \bibitem{elastic}
 S. Simayi,
  K. Sakano,
    H. Takezawa,
   M. Nakamura,
     Y. Nakanishi,   
           K. Kihou,   
                      M. Nakajima,   
  C. Lee,   
     A. Iyo,   
        H. Eisaki,   
                 S. Uchida,   
M. Yoshizawa,   
  J. Phys. Soc. Jpn
  \textbf{82}, 114604 (2013).

\end{thebibliography}
\end{document}